\documentclass[aps,prx,reprint,superscriptaddress,nofootinbib,longbibliography]{revtex4-2}

\usepackage[utf8]{inputenc}
\usepackage[T1]{fontenc}
\usepackage{lmodern}
\usepackage{array}
\usepackage{microtype}
\usepackage{mathtools}

\usepackage{placeins}
\usepackage{xspace}
\usepackage{multirow}
\usepackage{rotating}
\usepackage{caption}
\usepackage{subcaption}
\usepackage{tabularx}

\usepackage{amsmath}
\usepackage{amsfonts}
\usepackage{graphicx}
\usepackage{hyperref}
\hypersetup{
  colorlinks=true,
  citecolor=blue,
  linkcolor=blue,
  urlcolor=blue
}

\usepackage{mdframed}
\usepackage{comment}
\usepackage[braket,qm]{qcircuit}
\usepackage{multirow}

\newcommand{\cnot}{\textsc{cnot}}
\newcommand{\fiim}{\textsc{fiim}}
\newcommand{\riim}{\textsc{riim}}
\newcommand{\siim}{\textsc{siim}}
\newcommand{\liim}{\textsc{liim}}

\providecommand{\CNOT}{\ensuremath{\mathrm{CNOT}}\xspace}

\newcolumntype{P}[1]{>{\centering\arraybackslash}p{#1}}

\begin{document}

\title{Computationally Efficient Zero Noise \\ Extrapolation for Quantum Gate Error Mitigation}

\author{Vincent R. Pascuzzi\textsuperscript{\#}}\thanks{Work done while at Lawrence Berkeley National Laboratory}
\email{pascuzzi@bnl.gov}
\affiliation{Computational Science Initiative, Brookhaven National Laboratory, Upton, NY 11972, USA}
\author{Andre He\textsuperscript{\#}}
\email{andrehe@lbl.gov}
\affiliation{Physics Division, Lawrence Berkeley National Laboratory, Berkeley, CA 94720, USA}
\author{Christian W. Bauer}
\email{cwbauer@lbl.gov}
\author{Wibe A. de Jong}
\email{wadejong@lbl.gov}
\affiliation{Physics Division, Lawrence Berkeley National Laboratory, Berkeley, CA 94720, USA}
\author{Benjamin Nachman}
\email{bpnachman@lbl.gov}
\affiliation{Physics Division, Lawrence Berkeley National Laboratory, Berkeley, CA 94720, USA}

\begin{abstract}
Zero noise extrapolation (ZNE) is a widely used technique for gate error mitigation on near term quantum computers
because it can be implemented in software and does not require knowledge of the quantum computer noise parameters.
Traditional ZNE requires a significant resource overhead in terms of quantum operations.
A recent proposal using a targeted (or random) instead of fixed identity insertion method (\textsc{riim} versus \textsc{fiim})
requires significantly fewer quantum gates for the same formal precision.  We start by showing that \textsc{riim} can allow
for ZNE to be deployed on deeper circuits than \textsc{fiim}, but requires many more measurements to achieve the same level of
statistical uncertainty.
We develop two extensions to \textsc{fiim} and \textsc{riim}.
The List Identity Insertion Method (\textsc{liim}) allows to mitigate the error from certain \textsc{cnot} gates, typically
those with the largest error.
Set Identity Insertion Method (\textsc{siim}) naturally interpolates between the measurement-efficient \textsc{fiim} and the
gate-efficient \textsc{riim}, allowing to trade off fewer \textsc{cnot} gates for more measurements.
Finally, we investigate a way to boost the number of measurements, namely to run ZNE in parallel, utilizing as many quantum
devices as are available.
We explore the performance of \textsc{riim} in a parallel setting where there is a non-trivial spread in noise across sets of
qubits within or across quantum computers.  
\end{abstract}

\date{\today}
\maketitle

\makeatletter
\def\@fnsymbol#1{\ensuremath{\ifcase#1\or \#\or \dagger\or \ddagger\or
   \mathsection\or \mathparagraph\or \|\or **\or \dagger\dagger
   \or \ddagger\ddagger\or \# \else\@ctrerr\fi}}
\makeatother
\renewcommand{\thefootnote}{\fnsymbol{footnote}}
\footnotetext[1]{These authors contributed equally.}

\makeatletter
\def\@fnsymbol#1{\ensuremath{\ifcase#1\or *\or \dagger\or \ddagger\or
   \mathsection\or \mathparagraph\or \|\or **\or \dagger\dagger
   \or \ddagger\ddagger\or \# \else\@ctrerr\fi}}
\makeatother
\section{Introduction}
\label{sec:intro}

Noisy intermediate-scale quantum (NISQ)~\cite{Preskill2018quantumcomputingin} computers are promising tools for performing certain calculations more efficiently than can be computed with classical computers. This may allow for the evaluation of currently intractable calculations.
A fundamental challenge facing NISQ computation is that there is significant noise in the instruction sets (gates) and state readout.
The ultimate computational performance (towards fault tolerance) will be achieved with quantum error correction (see e.g. Ref.~\cite{RevModPhys.87.307}).  However, quantum error correction typically requires a many-to-one physical-to-logical mapping of quantum bits (qubits) and small enough gate errors. Current NISQ devices do not allow for the implementation of fault tolerant algorithms.

\begin{figure}[h!]
\centering
\begin{mdframed}
\centering
\leavevmode
\large
\Qcircuit @C=0.5em @R=0.8em @!R{
&&\lstick{\ket{0}}  &\gate{U_1}  	           &  \ctrl{1}  &  \qw   & \ctrl{1} &  \meter \\
&&\lstick{\ket{0}}  & \qw & \targ      &  \gate{U_2} &\targ   &\meter \\
&&&&&\downarrow&&\\
}\\\vspace{5mm}
\centering
\leavevmode
\Qcircuit @C=0.5em @R=0.8em @!R{
&&&&&\dstick{2n_1+1}&&&&&&&&\dstick{2n_2+1}&&&&\\
 \lstick{\ket{0}}  &\gate{U_1}  	         \gategroup{2}{3}{3}{9}{.5em}{.} \gategroup{2}{11}{3}{17}{.5em}{.}  &  \ctrl{1} &\qw &&\cdots & && \ctrl{1} &  \qw   & \ctrl{1} &\qw & &\cdots & && \ctrl{1}  & \meter \\
\lstick{\ket{0}}  & \qw & \targ      &\qw &&\cdots & && \targ &  \gate{U_2} &\targ  &\qw & & \cdots && & \targ   &\meter 
}
\end{mdframed}
\caption{An illustration of identity insertion for a generic controlled unitary operation with two qubits.  The $U_i$ represent unitary matrices and the $n_i$ are non-negative integers.}
\label{fig:circuitillustration}
\end{figure}
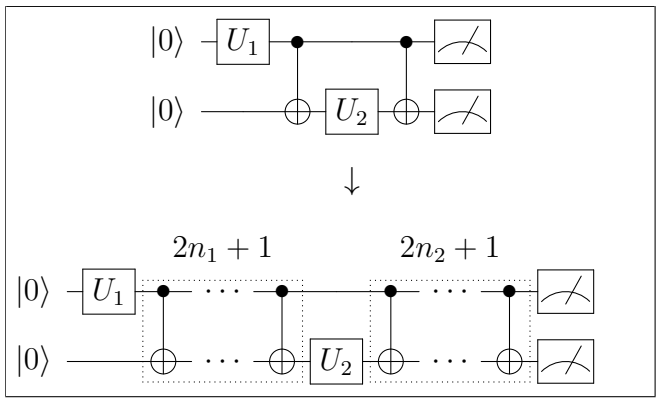

A variety of error mitigation strategies have been proposed on current and near-term quantum computers.
One widely used strategy is zero noise extrapolation (ZNE)~\cite{Richardson,Kandala_2019,Dumitrescu:2018,PhysRevX.8.031027,Temme_2017,Endo_2018,2003.04941,Giurgica_Tiron_2020,cai2021multiexponential}; additional approaches include estimation circuits~\cite{strikis2021learningbased,zlokapa2020deep,czarnik2021error,urbanek2021mitigating,Lowe_2021,czarnik2021qubitefficient}, quasi-probability methods~\cite{Temme_2017,PhysRevX.8.031027,Zhang_2020,mari2021extending}, and others -- see Ref.~\cite{Endo_2021} for a recent review.
In the ZNE protocol, measurements are made of an observable from a given circuit and a set of equivalent (auxiliary) circuits with amplified noise but the same zero-noise value. %
The noise is amplified in a controlled way so that measurements with different levels of noise can be used to extrapolate to the zero-noise limit. 
A hardware-agnostic approach to ZNE can be implemented by replacing a particular gate $U_i$ by $U_i(U_i^\dag U_i)^{n_i}$ for non-negative integer $n_i$ (see Fig.~\ref{fig:circuitillustration}).  
These $n_i$ identity insertions do not change the measured expectation value of the circuit, but since $U_i$ is noisy, the total error is increased.  
The standard Fixed Identity Insertion Method (\fiim)~\cite{Dumitrescu:2018,PhysRevX.8.031027} uses the same $n_i=n$ for every gate so that each gate is replaced by 
\begin{align}
    r = 2n+1
    \,,
\end{align}
copies of itself.  
Data are generated with $n=0,1,2...$ and then the target observable is extrapolated to $n=-\frac{1}{2}$, corresponding to $r = 0$.
This approach effectively mitigates gate errors, but at the expense of requiring a large number of quantum gates. 
ZNE is typically applied only to controlled-NOT (\cnot) gates which have a significantly higher error rate than single qubit gates\footnotetext{Future work can consider the application of this protocol to different sets of single- and multi-qubit basis gates.}.
For a circuit with $n_c$ \cnot\ gates, \fiim\ requires $2n\times n_c$ additional gates for each auxiliary circuit. 

A quantum gate efficient alternative ZNE called the Random Identity Insertion Method (\riim) was proposed recently~\cite{2003.04941}.  
Instead of using a global $n_i=n$, a non-uniform number of identity insertions is added for each gate.  
In particular, the first order correction is achieved by considering an input circuit having $n_c$ \cnot\ gates (and hence $n_c$ auxiliary circuits) where each auxiliary circuit has a different \cnot\ gate augmented with an identity insertion.  
Instead of requiring $2n_c$ gates for the lowest order correction as in \fiim, \riim\ requires only two additional gates for each auxiliary circuit.
In this way, \riim\ requires fewer gates for auxiliary circuits and so has the potential to enable gate error mitigation on deeper circuits than \fiim.
As we will discuss, \riim\ not only needs more circuits, it also necessitates the need for more measurements per circuit to achieve the same statistical uncertainty as \fiim. 

In this paper we develop methods to improve our ability to mitigate noise via ZNE. We develop variations of \fiim\ and \riim, and  study how one can parallelize the running of the required \riim\ circuits to obtain higher statistics. 
In particular, we propose new ZNE techniques called List and Set Identity Insertion Method (\liim\ and \siim) which replace only gates from a list (\liim) or sets of gates (\siim) with the same number of identity insertions instead of single gates (\riim) or all gates at once (\fiim).
We furthermore study a parallelization strategy for ZNE across quantum computers to generate a large number of measurements.  
In particular, we investigate the performance of \riim\ applied in parallel across computers with a non-trivial distribution of errors. 

This paper is organized as follows.  
Section~\ref{sec:riim-fiim-overview} briefly reviews \fiim\ and \riim, providing an explicit example where \riim\ achieves the same fidelity for a deeper circuit than \fiim. 
Then, Sec.~\ref{sec:Extension} introduces extensions of \riim\ and \fiim, including \siim\ and \liim.  The parallelization of ZNE is then studied in Sec.~\ref{sec:applications} for a synthetic distribution of noise and then in Sec.~\ref{sec:parallel-exec} for a realistic distribution of noise.  The paper ends with conclusions and outlook in Sec.~\ref{sec:conclusions}.

\section{{\mdseries\fiim} and {\mdseries\riim} overview}
\label{sec:riim-fiim-overview}
The basic idea of ZNE methods is to measure a given observable at varying levels of noise, and using the measured dependence on the noise to extrapolate to the expected noiseless value.
The dominant source of instruction-level noise in current digital quantum computers arises from the 2-qubit entangling \cnot\ gate, and the dominant noise channel is the 2-qubit depolarizing channel. In a two-qubit scenario, the depolarizing channel is given by the quantum operation acting on the system's density matrix, $\rho$:
\begin{align}
    {\cal E}(\rho) = (1-\epsilon) \rho + \frac{\epsilon}{4} I
\,,
\end{align}
where $I$ is the $2\times 2$ identity matrix and the noise parameter $\epsilon$ is of order a percent on current NISQ machines.
The action of a single noisy (depolarizing) \cnot\ gate on a general density matrix $\rho$ can therefore be written as
\begin{align}
\label{C_nominal}
    \cnot_{kl}[\rho] = (1-\epsilon) U_C^{(kl)} \rho U_C^{(kl)} + \frac{\epsilon}{4} \rho_{\not kl} \otimes I_{kl}
\,,\end{align}
where $\rho_{\not kl}$ represents the density matrix after tracing over the $k$ and $l$ qubits and $U_C$ is the unitary operator corresponding to the \cnot\ gate.  As the action of two \cnot\ gates gives the identity, the application of an odd number $r$ of \cnot\ gates to the same $kl$ qubits produces
\begin{align}\label{eq:depol}
    \cnot^r_{kl}[\rho] = (1-r\epsilon) U_C \rho U_C + \frac{r\epsilon}{4} \rho_{\not kl} \otimes I_{kl}+\mathcal{O}(\epsilon^2)
\,.\end{align}

Given Eq.~\eqref{eq:depol}, one can analyze the result of the action of a given quantum circuit $C$ containing $n_C$ \cnot\ gates and a universal depolarizing error rate $\epsilon$. This circuit creates a density matrix, which to first order in $\epsilon$ can be written as~\cite{2003.04941}
\begin{align}
\label{C_general}
    C[\rho] = (1-n_c \epsilon)\rho_{\rm ex} + \epsilon \sum_{i=1}^{n_c} \rho_i+\mathcal{O}(\epsilon^2)
\,,\end{align}
where $\rho_{\rm ex}$ is the density matrix that would be obtained from a noiseless circuit, and $\rho_i$ denotes the density matrix obtained if the $i^\text{th}$ \cnot\ gate in the circuit is replaced by the depolarizing channel.  
In other words, $\rho_i$ is constructed by replacing the 2-qubit system that the $i^\text{th}$ \cnot\ gate acts on with the completely mixed state $I/4$. 
Defining a circuit $C_{r_1, \ldots r_{n_c}}$, which replaces the $i^\text{th}$ $n_c$ \cnot\ gate by $r_i$ copies of the same \cnot\ gate, one can write the action of this circuit as 
\begin{align}
\label{leading_order_dep}
    C_{\{r_1, \ldots r_{n_c}\}}[\rho] = \left(1 - \epsilon \sum_{i=1}^{n_c} r_i \right)\rho_{\rm ex} + \epsilon \sum_{i=1}^{n_c} r_i \, \rho_i+\mathcal{O}(\epsilon^2)
\,.\end{align}

Given these expressions, one can now derive the expressions for the ZNE versions \fiim\ and \riim, as introduced in Ref.~\cite{2003.04941}. 
In \fiim, one constructs from the nominal circuit, $C_\text{nom}$, an auxiliary circuit, $C_\text{\fiim}$
in which each \cnot\ is replaced by $r_i = 3,~ \forall~i$ \cnot\ gates. Given Eqs.~\eqref{C_general} and \eqref{leading_order_dep}, one can show that 
\begin{align}
\label{fiim_combination}
    C_{\fiim}[\rho] \equiv \frac{3}{2} C[\rho] - \frac{1}{2} C_{\{3,3,...,3\}}[\rho] = \rho_{ex} + {\cal O}(\epsilon^2)
    \,.
\end{align}
The exact density matrix can therefore be obtained from a linear superposition of the nominal circuit and the auxiliary circuit, up to errors that are quadratic in the depolarizing noise parameter $\epsilon$.
In \riim, one replaces only a single \cnot\ gate by three \cnot\ gates, but then adds the $n_c$ possible density matrices. 
Symbolically, this can be expressed as
\begin{align}
\label{riim_combination}\nonumber
    C_{\riim}[\rho] &\equiv \frac{2+n_c}{2} C[\rho] - \frac{1}{2} \sum_{\sigma\in S_{n_c}} C_{\sigma \{3,1,...,1\}}[\rho] \\
    &= \rho_{ex} + {\cal O}(\epsilon^2)
    \,,
\end{align}
where
\begin{align}\nonumber
    \sum_{\sigma\in S_{n_c}}& C_{\sigma \{3,1,...,1\}}[\rho] \\
    &=C_{3, 1, \ldots 1}[\rho] + C_{1, 3, 1, \ldots 1}[\rho] + \ldots + C_{1, \ldots 1, 3}[\rho]
    \,,
\end{align}
for permutation matrices $\sigma\in S_{n_c}$.

The above analysis can be extended to higher orders in $\epsilon$; in fact, 
by computing the ${\cal O}(\epsilon^2)$ correction term one can show that \riim\ has a correction that is a factor of 3 smaller than that of \fiim~\cite{2003.04941}.  In what follows, we will mostly focus on first-order corrections in $\epsilon$.

\begin{figure}
    \centering
    \includegraphics[width=0.48\textwidth]{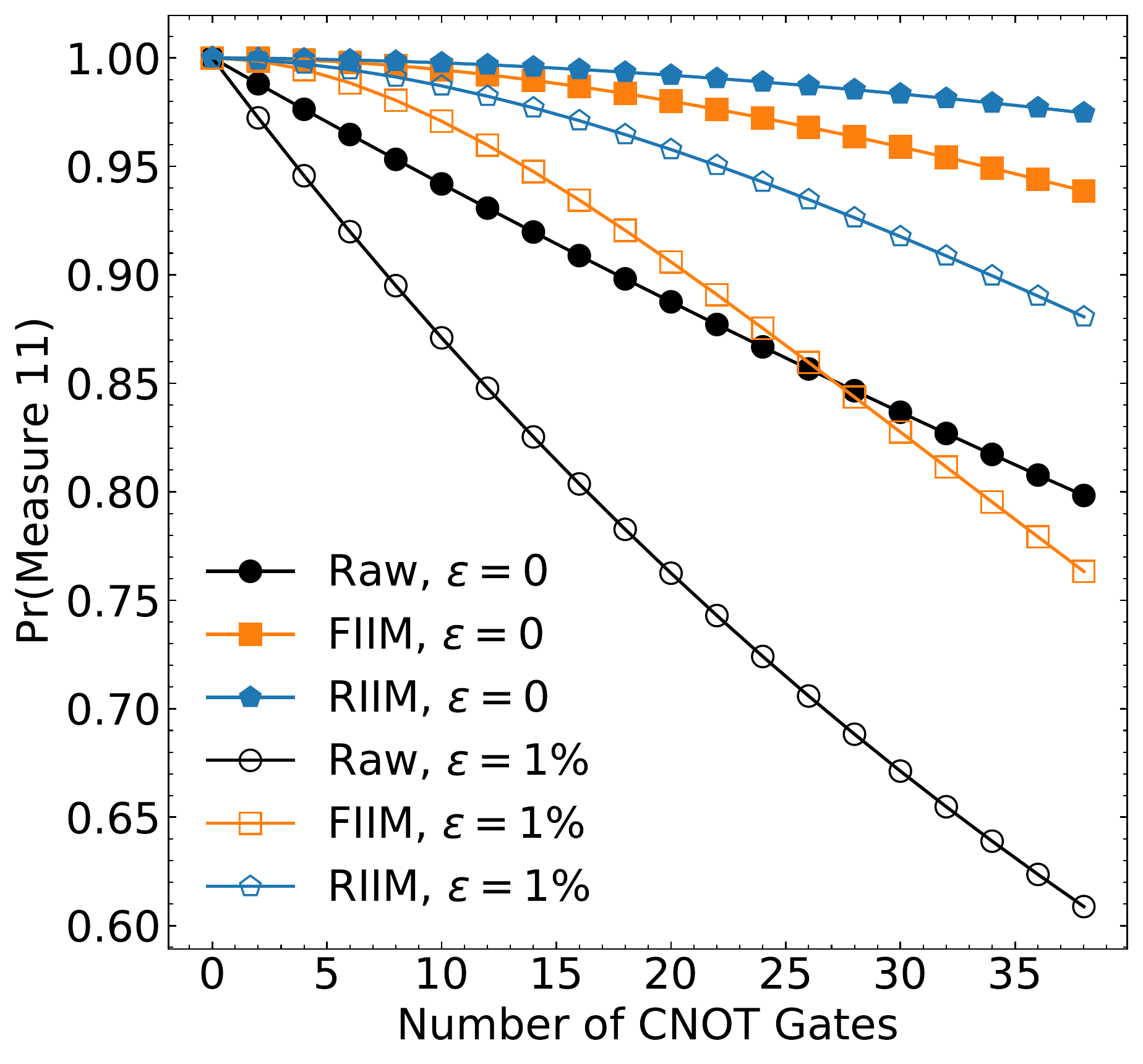}
    \caption{An illustration of the \fiim\ and \riim\ protocols in the absence of statistical noise for a two-qubit circuit with an even number of \cnot\ gates as specified by the horizontal axis.  The noise model includes depolarizing noise as specified in the legend and decoherence noise modeled by amplitude damping with a $T_1$ of 50 $\mu$s and a \cnot\ gate time of 200 ns.  The two qubits are prepared in the $\ket{1}$ state. Simulations performed with \textsc{Cirq}~\cite{cirq}. 
    }
    \label{fig:riimvfiim}
\end{figure}

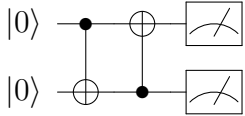
\begin{figure}[h!]
\centering
\leavevmode
\large
\Qcircuit @C=0.5em @R=0.8em @!R{
&&&\lstick{\ket{0}}  &  \ctrl{1}   &\qw  &\targ       & \qw& \meter \\
&&&\lstick{\ket{0}}  & \targ      &  \qw & \ctrl{-1} & \qw   &\meter \\
}
\caption{A simple two-qubit circuit comprising two \cnot\ gates followed by a measurement of each qubit.}
\label{fig:double}
\end{figure}

While \fiim\ and \riim\ both remove the linear depolarizing noise, they use different computational resources to achieve this goal.
\fiim\ requires a circuit that multiplies the total \cnot\ count by a factor of three, whereas \riim\ adds only two \cnot\ gates to its auxiliary circuit. 
This means that the \fiim\ auxiliary circuit is significantly deeper than in \riim\, and one can therefore expect \riim\ to outperform \fiim\, especially when the nominal circuit contains many \cnot\ gates.  
This is illustrated in Fig.~\ref{fig:riimvfiim} for a simple 2-qubit circuit (Fig.~\ref{fig:double}) with $2n+1$ \cnot\ gates in the absence of statistical noise.
Note that the simulation in Fig.~\ref{fig:riimvfiim} includes the effect of amplitude damping, for which $T_1=50~\mu$s, $T_\cnot{}=200$~ns, and the damping constant $\gamma = 1 - e^{-T_\cnot\//T_1}$; this is why the data for which $\epsilon=0$ are not unity.  In addition to mitigating depolarizing noise, ZNE also reduces the impact of amplitude damping.
%

On the other hand, \riim\ requires many more measurements to obtain the same statistical accuracy as \fiim. This can be seen from the linear combinations in Eqs.~\eqref{fiim_combination} and~\eqref{riim_combination} taken in \fiim\ and \riim.
Assume we know the value of $C[\rho]$ with statistical accuracy $\delta \rho_1$, while we know the value of each $C_{\{r_1, \ldots, r_{n_c}\}}[\rho]$ with accuracy $\delta \rho_2$. 
By taking the appropriate linear combinations, the standard deviation across measurements is given by
\begin{align}
    \sigma\left(C_{\fiim}[\rho]\right) & = \sqrt{\frac{9}{4} \delta \rho_1^2 + \frac{1}{4} \delta \rho_2^2}
 \nonumber\\
     \sigma\left(C_{\riim}[\rho]\right) & =  \sqrt{\frac{(1+2n_c)^2}{4} \delta \rho_1^2 + n_c \frac{1}{4} \delta \rho_2^2}
\,,
\end{align}
where the $n_c$ multiplying the $\delta\rho_2$ term in $C_{\riim}[\rho]$ arises from the fact that $\sum_{\sigma\in S_{n_c}} C_{\sigma \{3,1,...,1\}}[\rho]$ contains $n_c$ terms.
This implies that in order to get the same statistical precision in the two approaches requires a much more precise knowledge of $\delta\rho_1$ and $\delta \rho_2$ in \riim\ compared to \fiim. In particular, one requires
\begin{align}\label{eq:uncertscaling}
    [\delta \rho_1]_{\riim} \approx \frac{3}{2}\frac{[\delta \rho_1]_{\fiim}}{n_c} \,, \qquad [\delta \rho_2]_{\riim} = \frac{[\delta \rho_2]_{\fiim}}{\sqrt{n_c}}
\,.\end{align}
Given that the statistical error scales with the square of the number of measurements, Eq.~\eqref{eq:uncertscaling} shows that for \riim\ to match the \fiim\ precision, one needs $n_c^2$ more measurements for the nominal circuit and $n_c$ more measurements for each of each of the $C_{\{r_1, \ldots r_{n_c}\}}[\rho]$ circuits. 
Especially for large number of \cnot\ gates, for which \riim\ is especially expected to outperform \fiim, this is a potential drawback of \riim\ (but it is of course not an in principle limitation).
In the next section we will discuss a variant of \riim\, which allows for improvements over \fiim\ while keeping the number of measurements required more manageable. 
After that we discuss how one can parallelize the execution of the required quantum circuits, such that one can obtain the required number of measurements in a shorter amount of time. 

\section{Extending {\mdseries\fiim} and {\mdseries\riim} }
\label{sec:Extension}

In this section, we explore strategies for ZNE that use fewer quantum resources than \fiim\ and fewer measurements than \riim.

\subsection{Correcting for individual {\mdseries\cnot} noise: The List Identity Insertion Method ({\mdseries\liim})}
\label{sec:individual}

\begin{figure}
    \centering
    \includegraphics[width=0.48\textwidth]{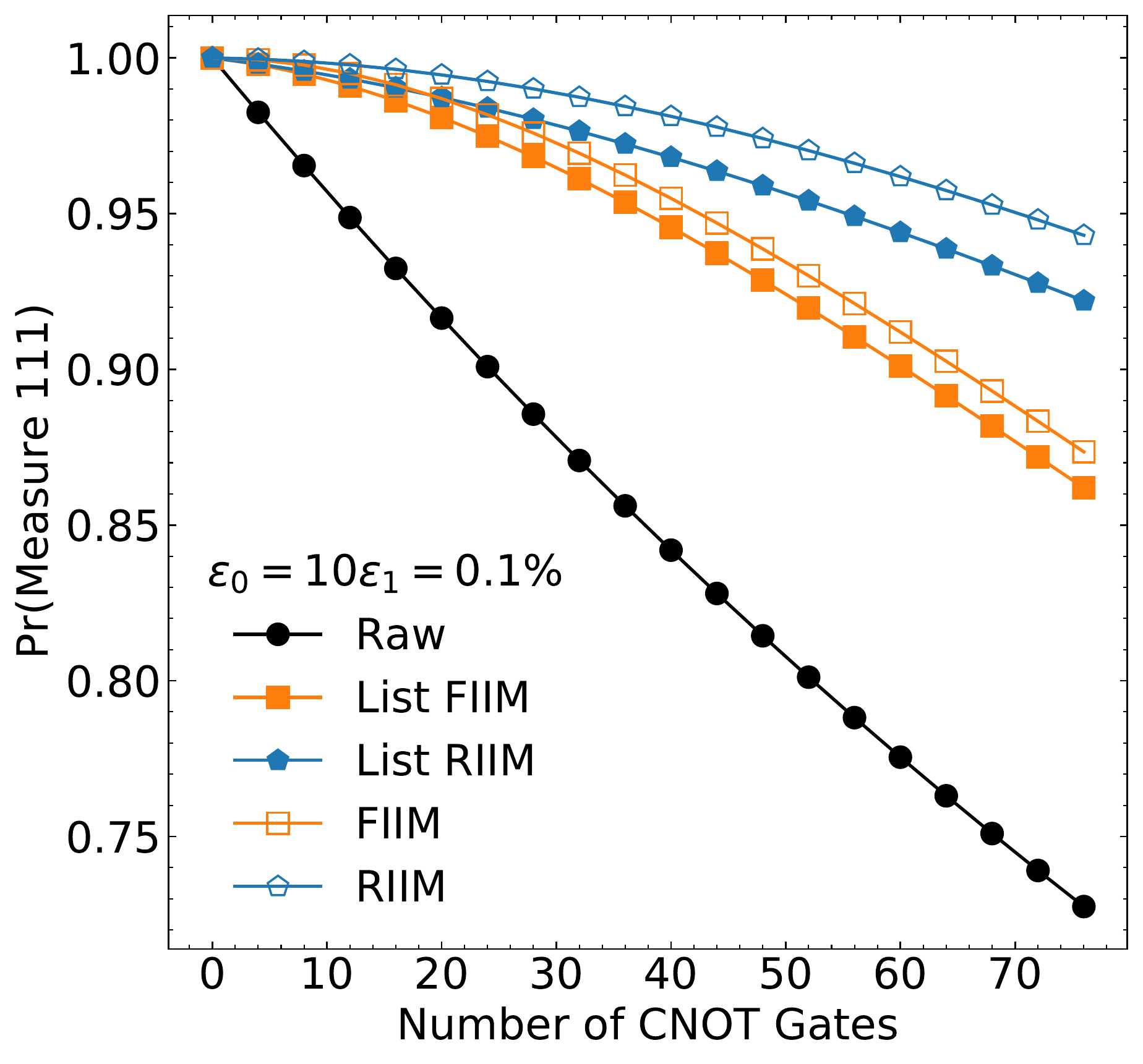}
    \caption{An illustration of the \liim\ protocol for the \fiim\ and \riim\ variants in the absence of statistical noise for a three-qubit circuit with an even number of \cnot\ gates between the first two and second two qubits as specified by the horizontal axis.  The noise model only includes depolarizing noise.  The three qubits are prepared in the $\ket{1}$ state. Simulations performed with \textsc{Cirq}~\cite{cirq}.}
    \label{fig:riimvfiim_list}
\end{figure}

In general, \cnot\ errors are different for each pair of qubits, and for systems in which some \cnot\ errors are much larger than the rest, one may be able to mitigate only the dominant \cnot\ errors with fewer resources than mitigating all errors.
Using a separate depolarizing noise value ($\epsilon_i$) for each \cnot\ gate, the leading order expression Eq.~\eqref{leading_order_dep} of the depolarizing noise becomes
\begin{align}\nonumber
        C_{\{r_1, \ldots r_{n_c}\}}&[\rho] =\\ &\left(1 - \sum_{i=1}^{n_c} \epsilon_i r_i \right)\rho_{\rm ex} + \sum_{i=1}^{n_c} \epsilon_i r_i \, \rho_i+\mathcal{O}(\epsilon_i\epsilon_j)
\,.
\end{align}
We can define a circuit $C_{\{3\}_{L}^{\rm all}}[\rho]$ that replaces all \cnot s of a given list $L$ by 3 \cnot s, and the sum of circuits $C_{\{3\}_{L}^{\rm sum}}[\rho]$ that replace one \cnot\ from each in this list by 3 \cnot s.
Then, one can construct a \fiim\ version that removes the linear terms of the large \cnot\ errors
\begin{align}\nonumber
        C_{\fiim, {L}}[\rho] &\equiv \frac{3}{2} C[\rho] - \frac{1}{2} C_{\{3\}_L^{\rm all}}[\rho] \\
        &= \rho_{ex} + {\cal O}(\{\epsilon_k\}_{k\not\in L}, \{\epsilon_i \epsilon_j\}_{i,j\in L})
    \,.
\end{align}
The correction remains linear in the $\epsilon$ terms that were not part of the list $L$, but those that were part of the list are multiplied by one other value of $\epsilon$. 
Similarly, one can obtain an analogous version of \riim:
\begin{align}\nonumber
        C_{\riim, {L}}[\rho] &\equiv \frac{2+|L|}{2} C[\rho] - \frac{1}{2} C_{\{3\}_L^{\rm sum}}[\rho] \\
        &= \rho_{ex} + {\cal O}(\{\epsilon_k\}_{k\not\in L}, \{\epsilon_i \epsilon_j\}_{i,j\in L})
    \,,
\end{align}
where $|L|$ denotes the number of elements in the list.  An illustration of the list-based ZNE is presented in Fig.~\ref{fig:riimvfiim_list}.

\subsection{Interpolating between {\mdseries\fiim} and {\mdseries\riim}: The Set Identity Insertion Method {\mdseries\siim}}
\label{sec:siim}

In the case that all of the \cnot\ errors are comparable and need to be mitigated, it is still possible to selectively replace gates by viewing \fiim\ and \riim\ as special cases of a more general approach, which we coin the Set Identity Insertion Method (\siim). 
In \siim\ one divides the $n_c$ \cnot\ gates into $n_s$ sets\footnote{We are assuming here that $n_c/n_s$ is an integer to simplify the notation.  The general approach still works if this is not true, but is a little more complicated to explain.} containing the same number of \cnot\ gates $m = n_c / n_s$. 
One is then free to choose some sets in $n_s$ to replace each \cnot\ gate by e.g. three \cnot{}, while keeping the other sets untouched.
We denote this by $C_{\{1\}, \ldots, \{3\}, \ldots \{1\}}[\rho]$.
Adding all different sets results in
\begin{align}
    C_{\{3\}^{m}}[\rho] \equiv C_{\{3\},  \{1\}, \ldots \{1\}}[\rho] + C_{\{1\}, \{3\}, \{1\}, \ldots \{1\}}[\rho] + \ldots 
\,,\end{align}
which contains a total of $n_s$ terms, with $m$ \cnot\ gates replaced.
Following the same steps as for \fiim\ and \riim\, one can now define the linear combination
\begin{align}
\label{siim_combination}
    C_{\siim}^{(n_s)}[\rho] \equiv \frac{2+n_s}{2} C[\rho] - \frac{1}{2} C_{\{3\}^{n_s}}[\rho] = \rho_{ex} + {\cal O}(\epsilon^2)
    \,.
\end{align}
Note that \fiim\ is recovered by using a single set $n_s = 1$, while \riim\ is obtained by using as many sets as \cnot\ gates $n_s = n_c$. 

\begin{figure}
    \centering
    \includegraphics[width=0.48\textwidth]{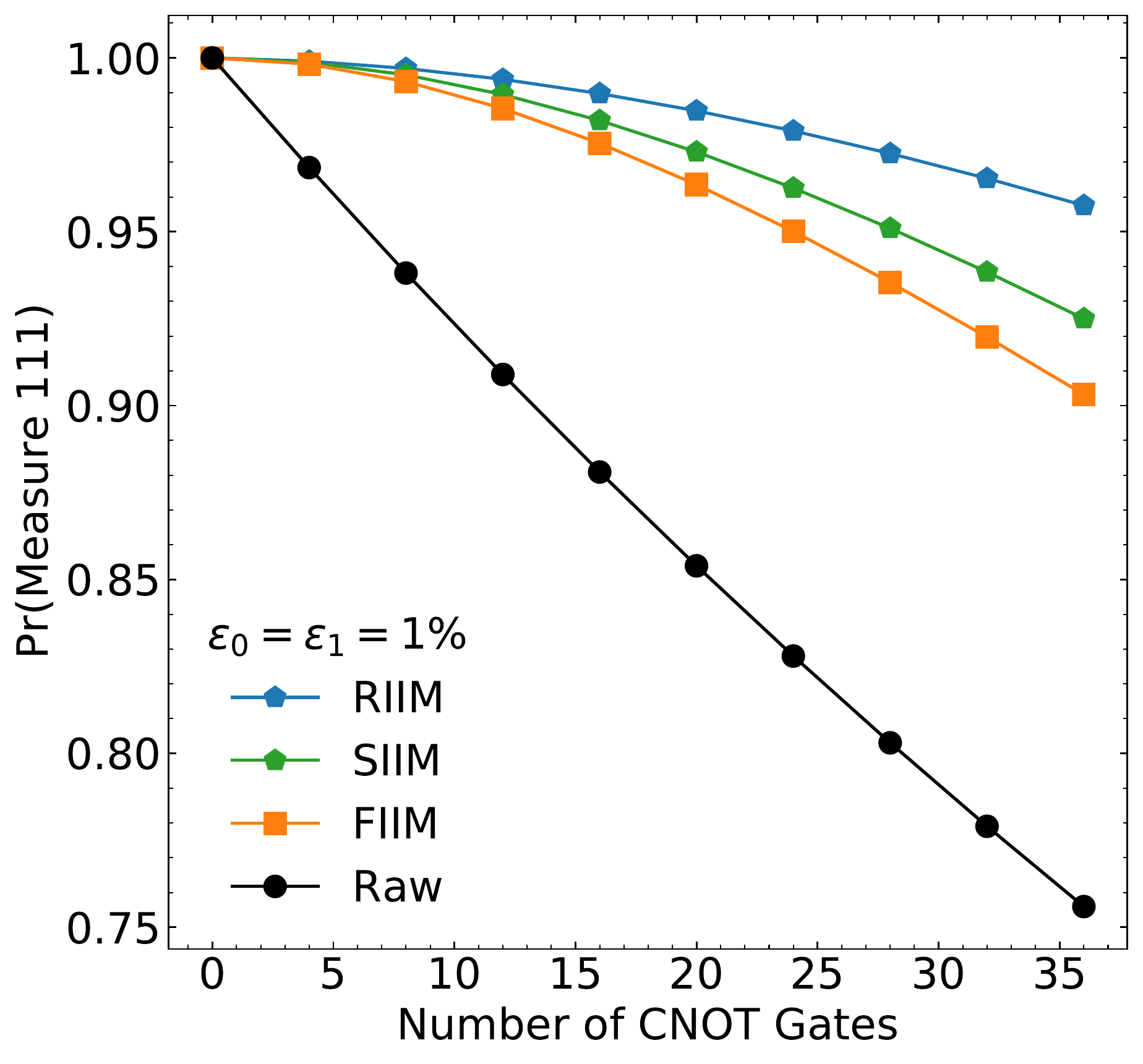}
    \caption{An illustration of the \siim\ approach (with two sets) in the absence of statistical noise for a three-qubit circuit with an even number of \cnot\ gates between the first two and second two qubits as specified by the horizontal axis.  The noise model only includes depolarizing noise.  The three qubits are prepared in the $\ket{1}$ state. Simulations performed with \textsc{Cirq}~\cite{cirq}.}
    \label{fig:riimvfiimvsiim}
\end{figure}

The \siim\ keeps much of the advantage of \riim, while greatly reducing its main disadvantage. In particular, the the extra number of gates required in \siim\ is $2 n_c / n_s$, while the extra number of measurements scales with $n_s^2$ instead of $n_c^2$ as for \riim.  Figure~\ref{fig:riimvfiimvsiim} illustrates how \siim\ interpolates between \riim\ and \fiim.
For purely depolarizing noise, the coefficient of the remaining $(n_c \epsilon)^2$ term in \siim\ is smaller by a factor of $(2+n_c/n_s) / 3 n_c$ compared to \fiim. This is summarized in the following table.
\begin{table}[h!]
\begin{tabular}{|c|c|c|c|c|}
\hline
Approach & \# of \cnot & error & \multicolumn{2}{c|}{\# of measurements}\\
& & (rel to \fiim) & Nominal & Correction \\\hline
\fiim & $3 n_c$ & 1 & $N_{\rm nom}$ & $N_{\rm corr}$\\
\riim & $n_c + 2$ & $\frac{1}{3}$ & $N_{\rm nom} \frac{(1+2n_c)^2}{9}$ & $N_{\rm corr}  n_c$  \\
\siim & $n_c + 2\frac{n_c}{n_s}$ &  $\frac{\delta}{1+2n_s}$ & $N_{\rm nom} \frac{(1+2n_s)^2}{9}$ & $N_{\rm corr}  n_s$  \\\hline
\end{tabular}
\caption{Summary of the three different ZNE approaches. As discussed in the text, the \riim\ approach requires almost a factor of three fewer \cnot\ gates to achieve a uncertainty that is smaller by a factor of 3. However, it does require many more events to reach the same statistical precision. The \siim\ approach interpolates between the two methods, allowing to reach a trade off between the number of \cnot\ gates and number of events needed.} 
\end{table}

\section{Paralleling ZNE: synthetic error models}
\label{sec:applications}

The results in the previous section focused on the case of zero statistical noise from a finite number of measurements. 
This is of course not realistic, and to reduce the statistical noise one will need to perform a large number of measurements.
One way to rapidly accumulate a large number of measurements is to parallelize across computers.  
This section will explore this idea first by considering a simple model in which depolarizing errors are normally distributed across a batch of quantum computers and then second with a realistic distribution of errors.

\subsection{Analytical Results}

Suppose that the depolarizing error is constant within quantum computer $i$ and that this value across computers is normally distributed with $\epsilon_i\sim\mathcal{N}(\epsilon,\sigma^2)$.  There is no unique way to parallelize the ZNE approaches introduced earlier.  One possibility is to run the nominal circuit and the noise-amplified auxiliary circuits on every computer and then average the results:

\begin{align}\nonumber
    \langle C_\text{ZNE}[\rho]\rangle &=\label{eq:cex} \rho_{ex}+\mathcal{O}(\langle\epsilon_i^2\rangle)\\
    &=\rho_{ex}+\mathcal{O}(\epsilon^2+\sigma^2)\,,
\end{align}
where we assume that $\epsilon\ll 1$ and $\sigma\lesssim\epsilon$ so that higher-order terms can be neglected.  Another strategy would be to run different parts of the ZNE calculation on different computers, where in general $\epsilon_{i} \neq \epsilon_{j}$. From Eqs.~\eqref{C_general} and \eqref{leading_order_dep}, this would lead to 

\begin{align}\nonumber
    \langle C_\text{ZNE}[\rho]\rangle &=\label{eq:cex2} \rho_{ex}+\mathcal{O}(\langle\epsilon_i-\epsilon_j\rangle)+\mathcal{O}(\langle\epsilon_i^2\rangle)\\
    &=\rho_{ex}+\mathcal{O}(\epsilon^2+\sigma^2)\,,
\end{align}
which is an equivalent expression to Eq.~\eqref{eq:cex}. However, executing the nominal and auxiliary circuits on every computer---as
opposed to spreading the computations across different computers---has the advantage of a smaller variance since the difference $\epsilon_i - \epsilon_j$ in Eq.~\eqref{eq:cex2} does not in general cancel.

\subsection{Numerical Results}

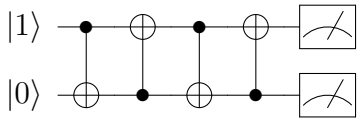
\begin{figure}[h!]
\centering
\leavevmode
\large
\Qcircuit @C=0.5em @R=0.8em @!R{
&&&\lstick{\ket{1}}  &  \ctrl{1}   &\qw  &\targ       & \qw& \ctrl{1}  &\qw&\targ       & \qw& \meter \\
&&&\lstick{\ket{0}}  & \targ      &  \qw & \ctrl{-1} & \qw   & \targ      &  \qw & \ctrl{-1} & \qw   &\meter \\
}
\caption{A simple two-qubit test circuit with four \cnot\ gates, followed by measurement of each qubit.}
\label{fig:fourgate}
\end{figure}

\begin{figure}[ht]
    \includegraphics[width=0.48\textwidth]{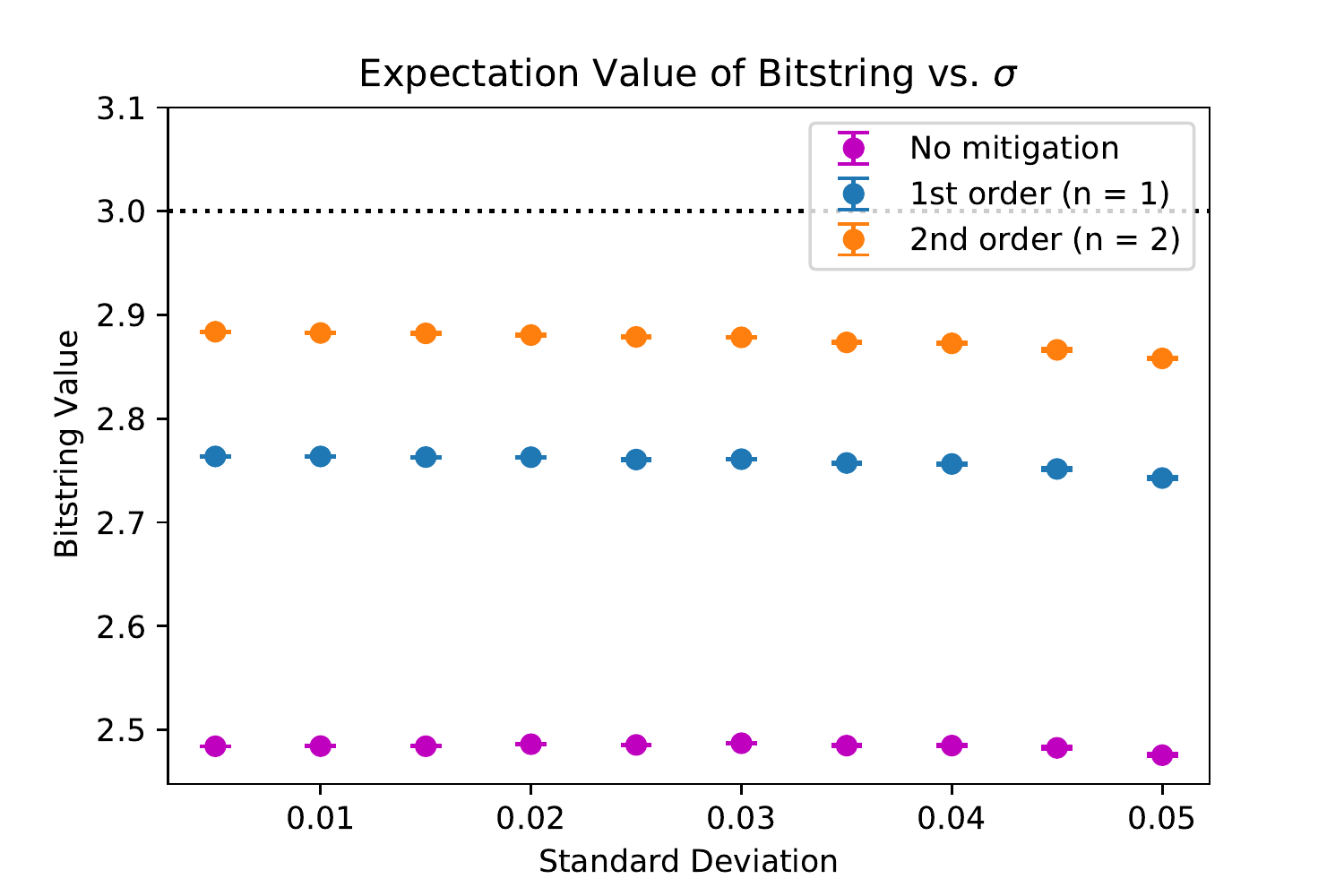}
\caption{ A demonstration of the scaling of extrapolation error in the 4-\CNOT\ circuit as the standard deviation of the gate error is increased. Simulations were run using Qiskit. A total of $10,000$ different error rates were sampled, and each instance of the circuit was run with $10,000$ shots.  
}
\label{fig:normal_ensemble}
\end{figure}

\begin{figure}[!ht]
    \includegraphics[width=0.48\textwidth]{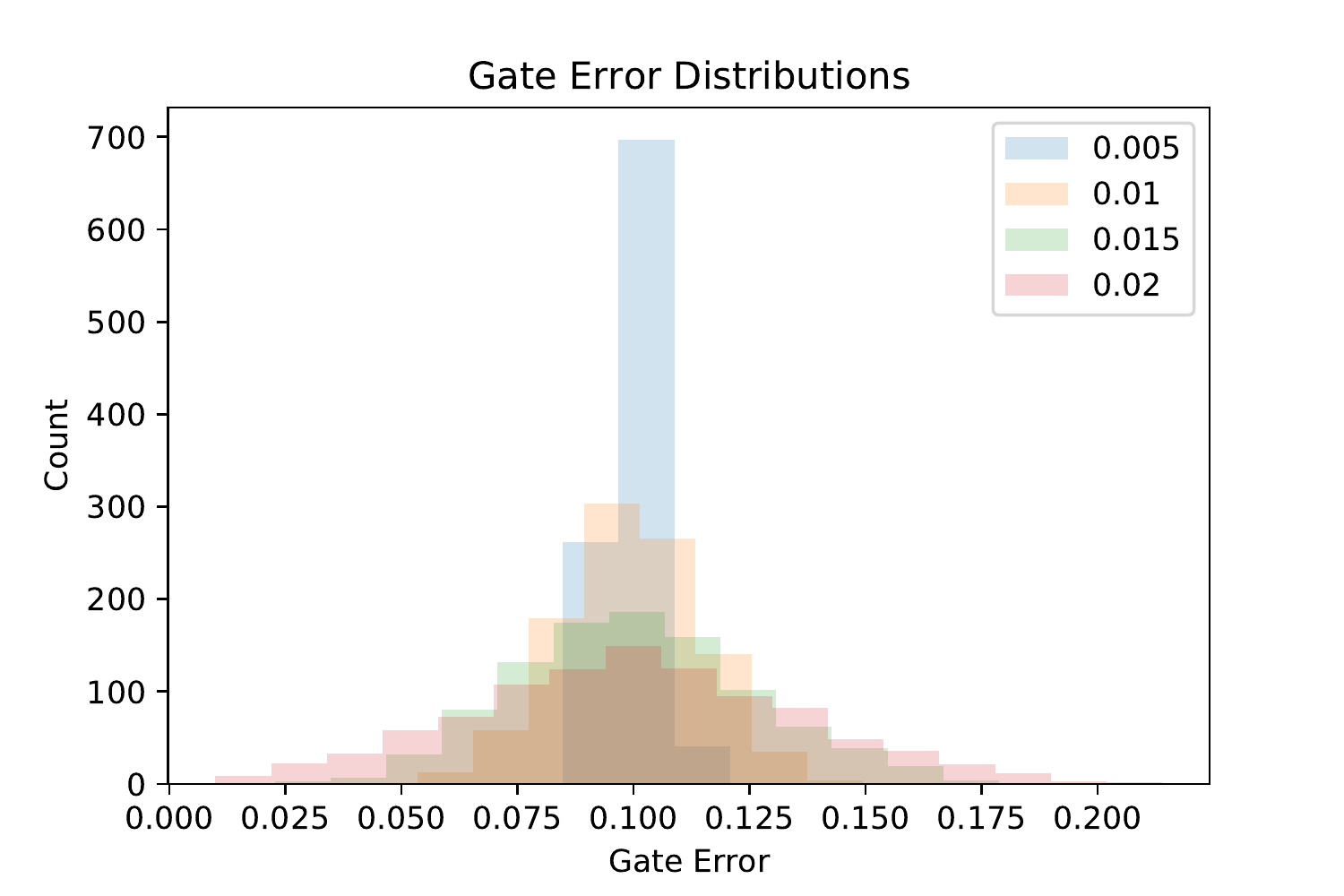}
\caption{Distribution of the error rates used to generate the data in Fig.~\ref{fig:normal_ensemble}.
}
\label{fig:normal_errs}
\end{figure}

As shown in the previous section, relaxing the assumption of uniform gate errors introduces an additional error term to the overall extrapolation error which, in the case of a normally-distributed set of gate errors, is dependent on the standard deviation of the distribution. 
In Fig.~\ref{fig:normal_ensemble}, the 4-\cnot\ circuit shown in Fig.~\ref{fig:fourgate} with a initial state \ket{10} 
was executed across an ensemble of simulated quantum devices in \textsc{Qiskit}~\cite{gadi_aleksandrowicz_2019_2562111}.  
When interpreting the output string as a integer, the ideal result in the absence of noise is 3 for a given measurement
The 2-qubit gate errors in each of these simulated devices are drawn from a normal distribution with $\mu = 0.1$, corresponding to a mean \CNOT\ error of $10\%$, while the standard deviation of the gate error distribution is increased to study the scaling of the error incurred by a wider distribution of errors.  
These errors are larger than typical uncertainties on existing machines, but the large mean ensures that the samples values are positive and to clearly demonstrate the scaling behavior.
The circuit was simulated across each of these devices, and the observable we use is the average value of the state, interpreting the bitstring in binary\footnote{Note that while this observable has been studied in other contexts, it has the feature that averaging integer values can artificially enhance the apparent fidelity when migrations in opposite directions cancel.}.  These are averaged across devices. 
Results from non-error-mitigated circuits are included, as well as results from error-mitigating using first and second order \textsc{FIIM}. 
The additional error is the excess error induced by the $\mathcal{O}(\epsilon^2+\sigma^2)$ term over the $\mathcal{O}(\epsilon^2)$ extrapolation error seen when assuming a single error rate across all devices and gates. 
The experiments shown in Fig.~\ref{fig:normal_ensemble} indicate that although there is a slight increase in the additional error as $\sigma$ is increased, it remains fairly small in magnitude. 
Figure~\ref{fig:normal_errs} provides a visualization of the error distributions used to generate these simulated ensembles. 
\section{Paralleling ZNE: realistic error models}
\label{sec:parallel-exec}
We turn now to simulations using two-qubit depolarizing noise parameters, $\epsilon_i$, extracted from devices currently available through IBM Quantum (IBM Q).
At the time of writing, IBM Q ``advanced access'' offers 14 unique systems (excluding simulators) with various properties such as depolarizing error parameters, T1 and T2 relaxation time constants, and gate execution times.
As retrieved from IBM Q backend (\textit{system}) properties, depolarizing error parameters, $\epsilon_{i,(jk)}$, where $i$ enumerates the systems themselves and $(jk)$ refers to the coupling between qubits $j$ and $k$, are symmetric about the qubits;
that is, $\epsilon_{i,(jk)} = \epsilon_{i,(kj)}$.
In the following studies, we consider depolarizing error channels in simulations performed using Qiskit to evaluate the utility of parallelization across multiple systems.

In the case of \textsc{RIIM}, where the number of required measurements (shots), $n_\text{shots}$, is proportional to the square of the
number of \cnot\ gates in the circuit, error mitigation can be computationally time-consuming;
this applies to both simulation and execution on real systems.
Moreover, significant latency can be incurred if one desires the highest fidelity system available to execute their circuits, as cloud-based
systems typically use a first in, first out queuing mechanism unless the preferred system is otherwise reserved.
Additionally, current IBM cloud quantum systems have limitations on the number of shots per circuit -- typically 8192 -- and the total
number of circuits -- ranging from 75 to 900 -- a single job submission can contain.
Since extensive allocations to the ideal system may not always be available for a given experiment, spreading large workloads across multiple (manifestly error-prone) systems is beneficial in terms of higher throughput and reliability of measurements.

For our experiment, we consider the same circuit from the previous section (Fig.~\ref{fig:fourgate}).
One may artificially deepen the four-\cnot\ circuit to gain some insight into the performance of a given error mitigation technique by, for example, replacing each of the four \cnot\ gates with an odd integer number of them.
The application of \riim\ would result in $n_c$ auxiliary circuits which in practice could be greater than the circuit limit on a
cloud-based quantum system. 
Furthermore, the required number of measurements to perform the protocol could far exceed the per-circuit shot limit of a system.
This motivates the use of multiple systems for parallelizing the error mitigation method.

\begin{figure}[h!]
    \includegraphics[width=0.48\textwidth]{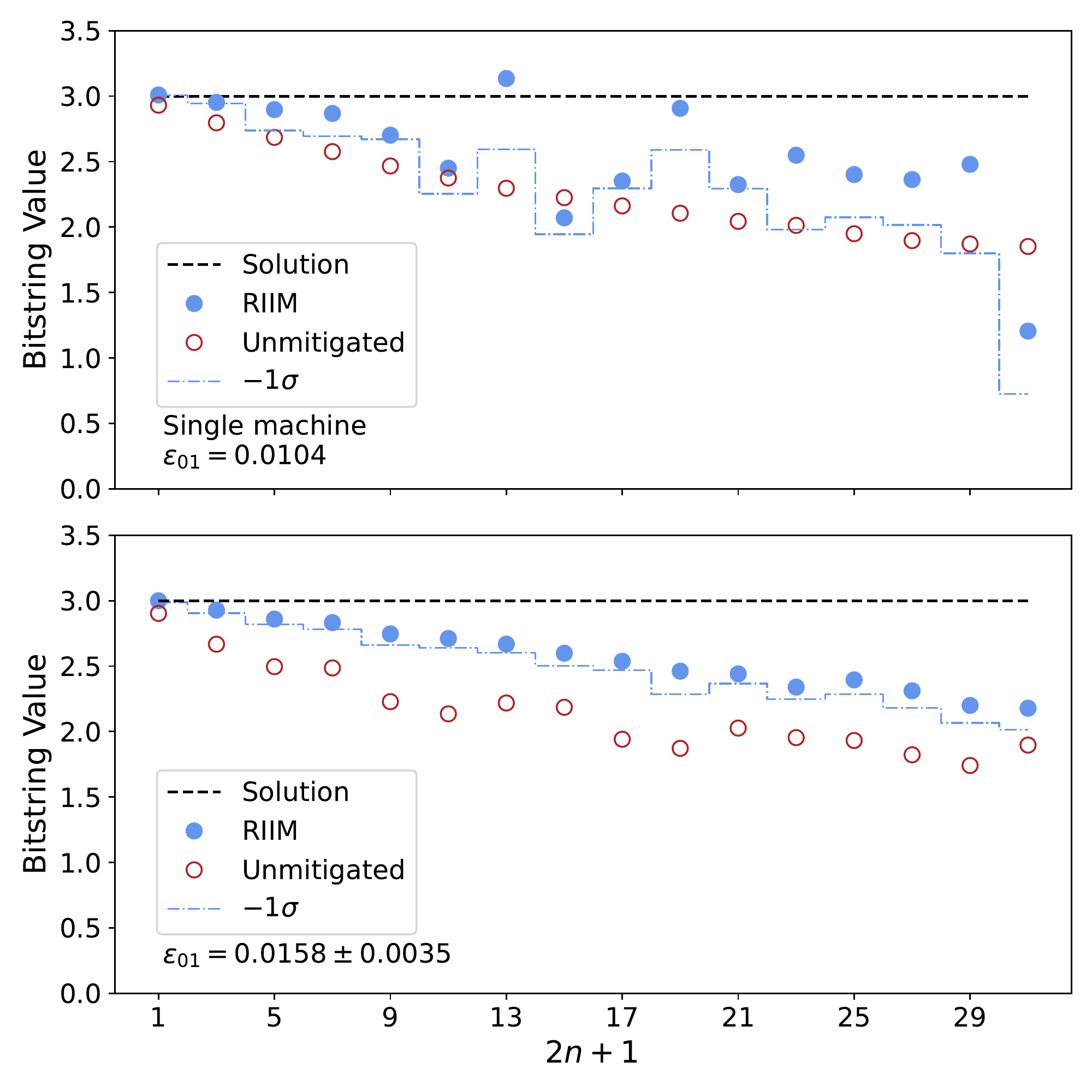}
\caption{Simulation results of the 4-\cnot{} circuit using a single-device (top) and multiple devices (bottom) with depolarizing gate noise.
The top figure consists of a single machine executing each circuit 8192 times, and the bottom an arbitrary set of
$4 \leq n_\text{NM} \leq 14$ noise models executing each circuit $8192\,n_\text{NM}$ times.
The gate noise in both cases is taken from IBM Q backend properties data.
}
\label{fig:riim_parallel_depol}
\end{figure}

\begin{figure}[t!]
    \includegraphics[width=0.48\textwidth]{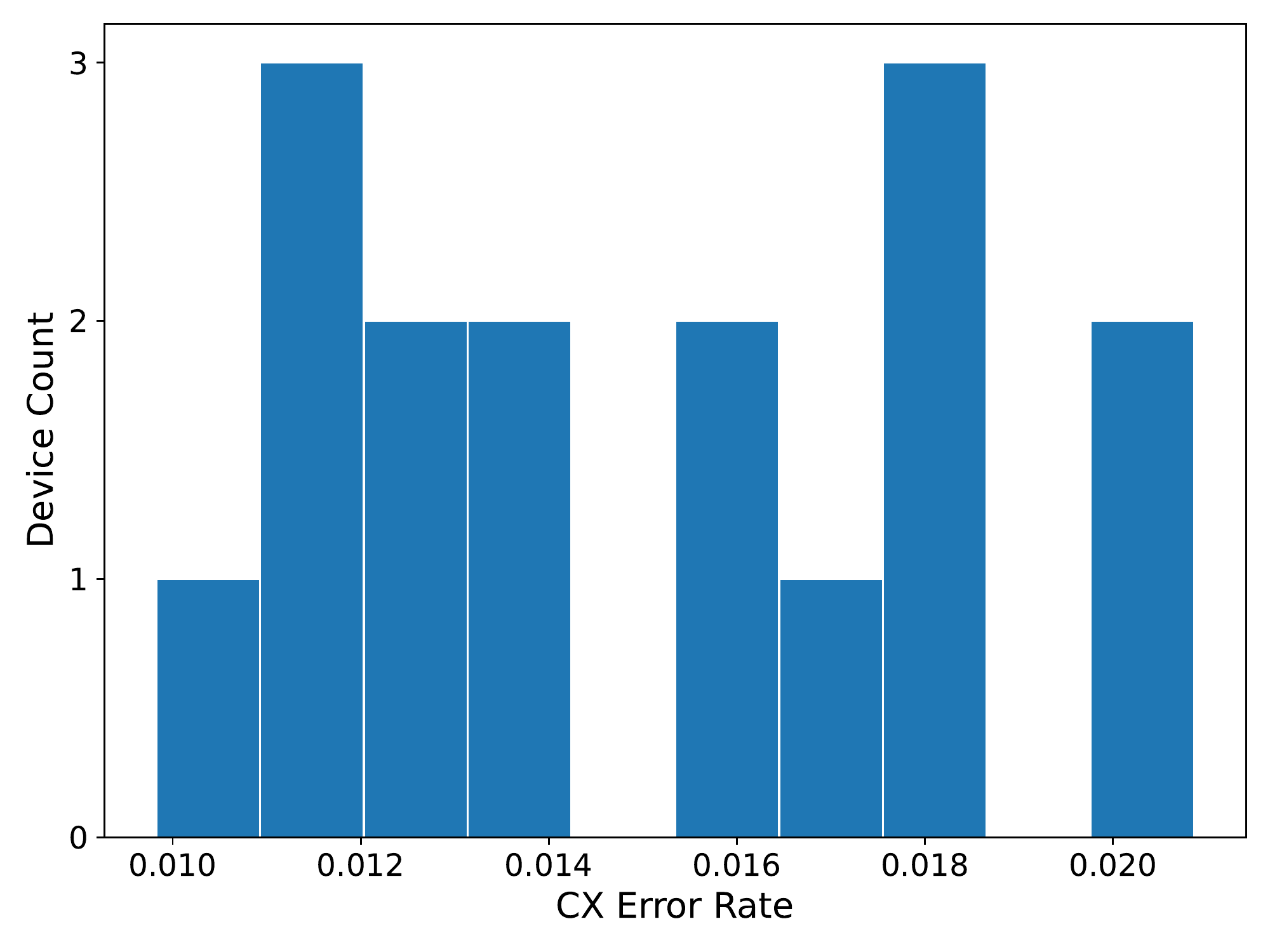}
\caption{Distribution of \cnot{} error rates applied in noisy simulations of the
4-\cnot{} circuit. Bins are centered at the error rate.
Data obtained from IBM Q backend properties.
}
\label{fig:cx_rates}
\end{figure}

Shown in Fig.~\ref{fig:riim_parallel_depol} are simulation results of executing the previously described circuit at various depths, ranging between 4--124 total \cnot\ gates, with only depolarizing noise applied; thermal relaxation and readout errors are disabled in
the simulations.
The observable measured, made in the standard basis, is the classical bit string value of the final state.
In the top panel are the unmitigated and \riim\ mitigated results using \texttt{ibmq\_guadalupe} with depolarizing error parameter $\epsilon_{01} = 0.0104$.
Each circuit was executed 8192 times.
The bottom panel shows the unmitigated and \riim\ output when executing the sets
of \riim\ circuits across multiple, arbitrarily chosen, systems.
The number of systems used in the calculation of each data point is equal to $2n+1$ ( \textit{i.e.}, the number of \cnot\ gates replacing a single \cnot\ in a circuit) with at most 14 unique systems (the maximum available from IBM Q at the time of writing) executing in parallel. 
For cases when $2n+1 > 14$, systems were selected -- again arbitrarily -- to be recycled and used for executing multiple sets of \riim\ circuits.
The mean depolarizing error parameter value of the 14 systems is $\epsilon_{01} = 0.0158 \pm 0.0035$, and includes the device used in the upper panel.
The distribution of machine \cnot\ error rates are shown in Fig.~\ref{fig:cx_rates}. 

The utility of parallelization permits larger numbers of measurements, executing a given set of circuits multiple
times, and thus greater efficacy when employing \riim.  
While the expected value in the lower panel cannot be better than the upper panel (the errors of the extra machines used for the additional data were all larger than the value of $\epsilon = 0.0104$ used in upper panel of Fig.~\ref{fig:riim_parallel_depol}), a significantly larger number of shots results in a much more stable result that often closer to the right answer than on the single best machine.
Additionally, the entire \riim\ protocol throughput increases, potentially reducing the time-to-solution by a factor of 14 (\textit{i.e.}, the number of available systems) on real hardware.
%
\section{Conclusions}
\label{sec:conclusions}
Zero noise extrapolation is a critical technique for error mitigation and is being used for a variety of experimental demonstrations on near term quantum devices (see e.g. Ref.~\cite{kim2021scalable,urbanek2021mitigating}).
While the core idea of ZNE is simple, there are many variations that can lead to improvements in resource usage and fidelity.
In this paper, we have introduced multiple approaches to ZNE.
First, we have developed a partial ZNE that applies to only a subset of qubits (List Identity Insertion Method).
Second, we showed that it is possible to generalize the Fixed and Random Identity Insertion Methods (\fiim\ and \riim) by replacing sets of qubits instead of all or single qubits.
This Set Identity Insertion Method (\siim) can trade-off the demanding gate resources of \fiim\ for the demanding measurement resources of \riim.
Lastly, we studied the parallelization of \riim\ in order to cope with the extensive measurement resources required to reach precision.

In this paper, we have focused on software and hardware agnostic methods for gate error mitigation.  Noise is amplified by means of identity insertions.  All of the methods introduced here could also be used with pulse-level gate lengthening for error amplification~\cite{Kandala:2019}.  Additionally, the ZNE methods introduced here can be combined with other gate error mitigation methods as well as readout error mitigation approaches~\cite{bialczak_quantum_2010,neeley_generation_2010,dewes_characterization_2012,magesan_machine_2015,debnath_demonstration_2016,song_10-qubit_2017,gong_genuine_2019,wei_verifying_2020,havlicek_supervised_2019,chen_detector_2019,chen_demonstration_2019,maciejewski_mitigation_2020,urbanek_quantum_2020,nachman_unfolding_2020,hamilton_error-mitigated_2019,karalekas_quantum-classical_2020,geller_efficient_2020,geller_rigorous_2020,2010.07496} for the ultimate error mitigation strategy.

\begin{acknowledgments}

We thank Aniruddha Bapat and Plato Deliyannis for detailed feedback on the manuscript.  This work is supported by the U.S. Department of Energy, Office of Science under contract DE-AC02-05CH11231. In particular, support comes from Quantum Information Science Enabled Discovery (QuantISED) for High Energy Physics (KA2401032) and the Office of Advanced Scientific Computing Research (ASCR) through the Accelerated Research for Quantum Computing Program.   This research used resources of the Oak Ridge Leadership Computing Facility, which is a DOE Office of Science User Facility supported under Contract DE-AC05-00OR22725.

\end{acknowledgments}

%
\bibliography{main}
\newpage
\appendix*
\onecolumngrid
\section{IBM Q Data and Code Availability}
Tables~\ref{tab:qubit-props} and~\ref{tab:gate-props} contain backend properties for each system used in this work.
Note that five of the fourteen systems have been retired since these studies were performed.
As such, we were unable to retrieve backend properties for the following systems: \texttt{ibmq\_16\_melbourne},
\texttt{ibmq\_athens}, \texttt{ibmq\_manhattan}, \texttt{ibmq\_paris} and \texttt{ibmq\_rome}.
\begin{table*}[!h]
        \setlength{\tabcolsep}{2pt}
        \begin{tabular}{ |c|c|c|c|c|c|c|c|c|c| } 
         \hline
         \multirow{2}{*}{System} & T1 [$\mu$s]  & T2 [$\mu$s]   & Frequency [GHz] & $P(0|1)$    & $P(1|0)$ \\
                                 & (Q0, Q1) & (Q0, Q1)  & (Q0, Q1)        & (Q0, Q1)    & (Q0, Q1) \\
         \hline
         \texttt{ibmq\_belem}       & 1.144[2], 9.928[1] & 1.144[2], 9.928[1] & 5.090, 5.246 & 2.720[-2], 3.660[-2] & 6.400[-3], 8.000[-3] \\
         \texttt{ibmq\_bogota}      & 8.777[1], 8.468[1] & 8.777[1], 8.468[1] & 5.000, 4.850 & 2.220[-2], 8.040[-2] & 1.000[-2], 2.040[-2] \\
         \texttt{ibmq\_casablanca}  & 1.202[2], 1.029[2] & 1.202[2], 1.029[2] & 4.822, 4.760 & 1.010[-1], 3.440[-2] & 1.560[-2], 8.600[-3] \\
         \texttt{ibmq\_guadalupe}   & 1.258[2], 1.097[2] & 1.258[2], 1.097[2] & 5.113, 5.161 & 1.820[-2], 2.140[-2] & 5.800[-3], 5.400[-3] \\
         \texttt{ibmq\_lima}        & 9.090[1], 9.864[1] & 9.090[1], 9.864[1] & 5.030, 5.128 & 3.480[-2], 4.460[-2] & 8.400[-3], 8.600[-3] \\
         \texttt{ibmq\_manila}      & 1.847[2], 1.396[2] & 1.847[2], 1.396[2] & 4.963, 4.838 & 3.160[-2], 3.060[-2] & 1.140[-2], 1.400[-2] \\
         \texttt{ibmq\_montreal}    & 1.019[2], 1.017[2] & 1.019[2], 1.017[2] & 4.911, 4.835 & 1.160[-2], 2.020[-2] & 6.800[-3], 8.000[-3] \\
         \texttt{ibmq\_quito}       & 7.571[1], 9.799[1] & 7.571[1], 9.799[1] & 5.301, 5.081 & 6.300[-2], 3.360[-2] & 1.820[-2], 1.140[-2] \\
         \texttt{ibmq\_santiago}    & 9.181[1], 7.148[1] & 9.181[1], 7.148[1] & 4.833, 4.624 & 2.940[-2], 1.460[-2] & 6.800[-3], 1.020[-2] \\
         \hline
        \end{tabular}
        \subcaption{Qubit properties}
        \label{tab:qubit-props}
        \bigskip
        \begin{tabular}{ |c|c|c|c|c|c|c|c|c|c| } 
         \hline
         \multirow{2}{*}{System}    & $X$-Gate Error        &  $X$-Gate Length [ns]            &  $CX$-Gate Error                         & $CX$-Gate Length [ns]       \\
                                    & (Q0, Q1)              &  ($X_{(0)}=X_{(1)}$)             &  ($\epsilon_{(0,1)}=\epsilon_{(1,0)}$)   & ($CX_{(0,1)}=CX_{(1,0)}$)   \\
         \hline
         \texttt{ibmq\_belem}       & 1.975[-4], 2.538[-4]  & 3.556[1]                        &1.801[-2]                                  & 8.107[2] \\
         \texttt{ibmq\_bogota}      & 1.947[-4], 2.464[-4]  & 3.556[1]                        &1.334[-2]                                  & 6.898[2] \\
         \texttt{ibmq\_casablanca}  & 2.672[-4], 2.804[-4]  & 3.556[1]                        &1.885[-2]                                  & 7.609[2] \\
         \texttt{ibmq\_guadalupe}   & 2.104[-4], 3.436[-4]  & 3.556[1]                        &1.437[-2]                                  & 3.342[2] \\
         \texttt{ibmq\_lima}        & 2.508[-4], 1.870[-4]  & 3.556[1]                        &1.172[-2]                                  & 3.058[2] \\
         \texttt{ibmq\_manila}      & 1.658[-4], 2.491[-4]  & 3.556[1]                        &2.120[-2]                                  & 2.773[2] \\
         \texttt{ibmq\_montreal}    & 1.835[-4], 1.605[-4]  & 3.556[1]                        &1.473[-2]                                  & 3.840[2] \\
         \texttt{ibmq\_quito}       & 6.078[-4], 2.839[-4]  & 3.556[1]                        &1.194[-2]                                  & 2.347[2] \\
         \texttt{ibmq\_santiago}    & 4.634[-4], 2.028[-4]  & 3.556[1]                        &1.838[-2]                                  & 5.262[2] \\
         \hline
        \end{tabular}
        \subcaption{Gate properties}
        \label{tab:gate-props}
\caption*{TABLE II: Backend properties of the systems used in the simulations presented in Sec.~\ref{sec:parallel-exec}.
These data correspond to the calibration data at the time our experiments were executed.
Numerals in square brackets represent power of 10.
Five systems---\texttt{ibmq\_16\_melbourne}, \texttt{ibmq\_athens}, \texttt{ibmq\_manhattan}, \texttt{ibmq\_paris} and \texttt{ibmq\_rome}---have since been retired and therefore are not included in this table.}
\label{app:ibmq-data}
\end{table*}

The source codes implementing the algorithms described in this paper will be available at:
\href{https://github.com/vrpascuzzi/computationally-efficient-zne}{https://github.com/vrpascuzzi/computationally-efficient-zne}.
\end{document}